\documentclass[prb,showpacs,twocolumn,floatfix]{revtex4}
\usepackage[english]{babel}
\usepackage{graphicx}
\usepackage{amsmath}
\usepackage{amssymb}

\begin{document}

\newcommand{\Tr}{{\rm Tr}}

\title{Generalization of the Poisson kernel to the superconducting random-matrix ensembles}

\author{B. B{\'e}ri}
\affiliation{Instituut-Lorentz, Universiteit Leiden, P.O. Box 9506,
2300 RA Leiden, The Netherlands}
\date{February 2009}

\begin{abstract}
 We calculate the distribution of the scattering matrix at the Fermi level for
chaotic normal-superconducting systems for the case of arbitrary coupling of the scattering
region to the scattering channels. 
The derivation is based on the assumption of uniformly distributed
scattering matrices at ideal coupling, which holds in the
absence of a gap in the quasiparticle excitation spectrum. The resulting distribution
generalizes the Poisson kernel to the nonstandard symmetry classes introduced
by Altland and Zirnbauer. We show that unlike the Poisson kernel, our result
cannot be obtained by combining the maximum entropy principle with the
analyticity-ergodicity constraint. As a simple application, we calculate the
distribution of the conductance for a single-channel chaotic Andreev quantum dot in a
magnetic field. 
\end{abstract}

\maketitle

\section{Introduction}
\label{sec:intro}
Statistical aspects  of electronic transport through chaotic cavities
(quantum dots) can be efficiently described  using a random matrix model for the $N\times
N$ unitary scattering matrix $S$ of the system (see Ref.~\onlinecite{RMTQTR} for a
review). For sufficiently low temperatures and voltages, transport properties
can be expressed by the scattering matrix at the Fermi level. Besides
unitarity, a crucial role is played in the random matrix models by 
the additional constraints satisfied by $S$, defining the so-called symmetry classes.

In the absence of superconductivity, following Dyson,\cite{dyson1962ste} three
symmetry classes are distinguished,
depending on the presence or absence of time-reversal and spin-rotation
symmetry. In this classification scheme,
the cases are labeled by the index $\beta$, and the additional constraints on $S$
are as follows.  In the presence of time-reversal, as well as spin-rotation
symmetry ($\beta=1$), $S$ is symmetric, $S=S^T$. In the
absence of time-reversal symmetry ($\beta=2$), the only requirement is the
unitarity of $S$. In the presence of time-reversal symmetry, but without
spin-rotation invariance ($\beta=4$), $S$ is self-dual, $S=S^R$. (The dual of
a matrix $A$ is defined by $A^R=\tau A^T \tau^T$, with $\tau=i\sigma_2$, where
$\sigma_j$   denotes the $j$-th Pauli matrix in spin space.)
It has been shown\cite{Brouw95gce} that for ideal coupling of the scattering channels to
the cavity, i.e., in the absence of direct reflection from the cavity openings,
the distribution of $S$ is uniform. Uniformity is understood with respect to the
invariant measure  in the unitary group subject to the constraints
imposed  by the symmetries under spin-rotation and time-reversal. 
From the uniform distribution at ideal coupling, it follows\cite{Brouw95gce} that at
arbitrary coupling the probability density of $S$ is given by  the Poisson kernel\cite{Brouw95gce,hua1963haf,krieger1965stc,mello1985ita,doron1992srd,baranger1996spa}
\begin{equation}
P_{\beta}(S)\propto |\det(1-r^\dagger S)|^{-(\beta N+2-\beta)}
\label{eq:PoissWD}\end{equation}
where $r$ is the matrix describing the direct reflections from the openings.

Dyson's classification scheme becomes insufficient in the presence
of superconductivity.\cite{Alt96,Alt97} In normal-superconducting hybrid systems, 
the scattering matrix acquires an electron-hole structure,\cite{takane1991,lambert1991glf,bee92a}
and it satisfies\cite{Alt97} a constraint at the Fermi level, $S=\Sigma_1 S^*
\Sigma_1$, expressing the electron-hole symmetry. 
($\Sigma_j$ denotes the $j$-th Pauli matrix in electron hole space.)
Altland and Zirnbauer\cite{Alt97} showed that depending
on the symmetries under time-reversal and spin-rotation,
these systems fall into four new symmetry classes, which they labeled following Cartan's
notation of the corresponding symmetric spaces. 
Systems where both symmetries are broken, belong to class $D$. 
If only spin-rotation invariance is broken, class $D$III is realized. If only
time-reversal symmetry is broken, the system belongs to class $C$, and
finally, if all symmetries are present, the system belongs to class $C$I.
The requirements for $S$ following from time-reversal and
spin-rotation symmetry are the same as in the absence of superconductivity.
Assuming gapless quasiparticle excitations, Altland and Zirnbauer introduced a random scattering matrix
model for transport in chaotic normal-superconducting
systems, by adopting a uniform distribution for the scattering matrix.
This is appropriate for the case when
the coupling of the cavity to the transport channels is ideal. 
The analogue of the Poisson kernel, i.e., the  distribution of $S$,  for the case of arbitrary coupling,
to the best of our knowledge, has not been presented yet. 
In this study, we aim at providing this result.
We believe that the knowledge of this distribution is desirable, as it can serve as a starting point to extend results
that are based on the Poisson kernel $P_\beta(S)$, from Dyson's standard
symmetry classes to the classes of Altland and Zirnbauer. 
As a particular example we mention the study of dephasing in the framework of B\"uttiker's
dephasing lead model.\cite{buett86a,buett88a,marcus93,BarMel95} In this model,  in order
to account for dephasing mechanisms that occur uniformly in the quantum dot, the knowledge of the distribution of the
scattering matrix for nonideal coupling is essential.\cite{BroBee97volprobe}

The paper is organized as follows. 
In the next Section, we relate the attributes of the scattering matrix to
those of normal-superconducting quantum dots, and
we briefly discuss the conditions at which a random
scattering matrix description for the  transport in such systems
is adequate.
In Sec.~\ref{sec:manifolds}, we detail the 
properties of the manifolds in the space of $N\times N$ matrices defined by the
constraints on the scattering matrix corresponding to the symmetry classes of Altland and Zirnbauer.
In Sec.~\ref{sec:sdist}, we present the calculation of the
distribution $P(S)$ of the scattering matrix, based on the assumption that $S$
is uniformly distributed in the ideal coupling case. 
We illustrate the use of our result in Sec.~\ref{sec:appliC} on a simple but physically realistic
example, a single mode normal-superconducting quantum dot in magnetic field.
We conclude in
Sec.~\ref{sec:concl} by contrasting  $P(S)$ and the Poisson kernel
$P_\beta(S)$ regarding  the applicability of the analyticity-ergodicity
constraint of Ref.~\onlinecite{mello1985ita}.

\section{Physical realization of the scattering matrix ensembles}
\begin{figure}
\begin{center}
\includegraphics[height=6cm]{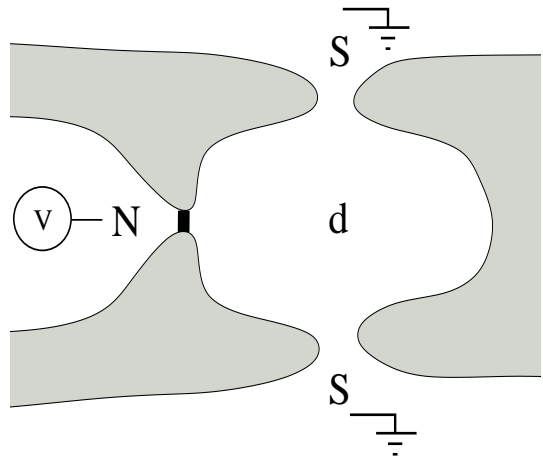}
\end{center}
\caption{An Andreev quantum dot formed by a normal conducting cavity (d) with two
  superconducting contacts (S).  In the transport state, an
  infinitesimal voltage $V$ is applied between the grounded superconductors and the
  normal reservoir (N) contacted at the left opening. 
The distribution of $S$ derived in this paper considers the
  effect of a tunnel barrier in the normal contact, indicated by a black
  rectangle in the figure.
   }
\label{fig:sketch}
\end{figure}

In the case of the symmetry classes of Altland and Zirnbauer, the role of the chaotic cavity is played by
a so-called chaotic Andreev quantum dot,\cite{Alt96,Alt97,Bee05}
i.e., a structure formed by a chaotic normal conducting quantum dot contacted to
superconductors. 
In the vicinity of the  Fermi level, there are no propagating modes in the
superconductors. 
We consider the situation when the Andreev quantum dot is contacted to normal
reservoirs. The number of propagating modes in the contacts to the normal reservoirs (normal
contacts for short), including  electron-hole degrees of freedom,  
defines  the size of the scattering matrix $S$.  
We concentrate on the regime where transport properties can be expressed in
terms  of the  scattering matrix at
the Fermi energy: the temperature and the
voltages applied to the normal reservoirs are  assumed to be
much smaller than the energy scale corresponding to the escape rate from the normal region
and the gap of the superconductors. (The superconductors are assumed to be grounded.) 
A sketch of an Andreev quantum dot with two
superconducting and one normal contact  is shown in Fig.~\ref{fig:sketch}. 
(Charge transport can already take place using one normal contact,
due to the Andreev reflection at the superconducting interfaces.\cite{BTK82,takane1991,lambert1991glf,bee92a}) 

By slightly varying the shape of the Andreev quantum dot, one obtains an
ensemble of systems and, therefore, an ensemble of scattering matrices. We
discuss below the   conditions at which this ensemble can realize the 
random scattering matrix models discussed in this paper. 
The only parameters that  enter the scattering matrix distribution are the symmetries  of
$S$ and the properties of the normal contacts. 
This implies that the conductance of the superconducting contacts should be
much larger than of the normal contacts, otherwise transport properties would
be sensitive to the ratio of these conductances.\cite{Clerk00acci,Samuel02cdsa}
In addition, Frahm {\it et al}\cite{Fra96} has shown that for the effect of the superconductors
on the dynamics in the cavity to be considerable, the Andreev conductances
of  the superconducting contacts should be  much larger than unity.\cite{Fra96} 
For a random scattering matrix description of transport, it is important
that the quasiparticle excitations are gapless.
If the excitations were gapped, the normal contacts to the Andreev quantum dot
would effectively act as  normal-superconductor interfaces,\cite{Clerk00acci,Samuel02cdsa}
directly reflecting incoming quasiparticles, i.e., during transport, the
quasiparticles would not explore the chaotic cavity.
Gapless chaotic Andreev quantum dots that belong
to class $C$ and $D$ can be realized with one
superconducting contact already, using a time reversal breaking magnetic field to suppress the
proximity gap.\cite{Mel96,Alt96,Fra96,Mel97}   For classes $C$I and
$D$III, time reversal invariance requires the absence of magnetic fields in
the dot. The gap can be suppressed\cite{Alt97,Mel97} by using two
superconducting contacts, with a phase
difference $\pi$. 
The assumption of a uniformly distributed scattering matrix
corresponds to assuming that the coupling of the cavity to the transport channels is ideal, i.e., that the
normal contacts are without a tunnel barrier. (The contacts to the
superconductors can contain tunnel barriers, as long as they satisfy the
aforementioned requirements for their conductances.) In the remaining part of the paper, our
task is to generalize this uniform distribution to one that accounts for nonideal
normal contacts. It is worthwhile to note here that we do not rely on the
specific details of the barriers in the normal contacts, we only use that the
scattering matrix of the barriers satisfies the same symmetry requirements
as the scattering matrix of the system without the barriers. Our calculation is
therefore  equally valid for contacts to the normal reservoirs with tunnel barriers that do not mix
electrons and holes, and for barriers that mix electrons and holes. The latter
situation can occur if there is a 
region with an induced superconducting gap in the contact to a normal
reservoir, that the quasiparticles have to tunnel
through  to reach the (gapless) cavity region.

\section{Scattering matrix manifolds}
\label{sec:manifolds}
The scattering matrix can be considered as a point of a manifold ${\cal M}_X$ in the space of $N\times N$ matrices,
where $X$ refers to  the symmetry class  under consideration. 
The distribution of the scattering matrix is understood with respect to  the
invariant measure on ${\cal M}_X$.
We first state the symmetry properties of ${\cal M}_X$ following
Ref.~\onlinecite{Alt97}.  We then take a common
route,\cite{tracy1994irm,RMTQTR} and consider ${\cal M}_X$ 
as a Riemannian manifold, to give expressions for the invariant arclength
$ds_X^2=\Tr(dUdU^\dagger)$  and the corresponding  measure $d\mu_X(U)$ in an
infinitesimal neighborhood of  $U\in {\cal M}_X$. 
As usual,\cite{Mehta,tracy1994irm,guhr1998rmt} we  parametrize this infinitesimal
neighborhood with the help of  infinitesimal
matrices $\delta U_X$, with symmetry properties
dictated by those of ${\cal M}_X$, such that the measure is
simply the product of the independent matrix elements of $\delta U_X$.

For class
$D$, the manifold ${\cal M}_D$ is isomorphic to SO($N$), with
$U=\nu {\cal O} \nu^\dagger$, ${\cal O}\in {\rm SO}(N)$, and                                
\begin{equation}
\nu=\frac{1}{2}
\left(\begin{array}{cc}
1+i & 1-i \\
1-i & 1+i
\end{array}\right), \quad \nu^2=\Sigma_1. 
\end{equation} 
It might be worthwhile to note here, that solely from the unitarity of $U$ and
the symmetry $U=\Sigma_1 U^* \Sigma_1$ only $\det U=\pm 1$  follows. In Ref.~\onlinecite{Alt97}, the
manifold ${\cal  M}_D$ was identified through the exponentiation of the Bogoliubov-de Gennes
Hamiltonian, which leads to $\det\ U=1$ due to the mirror symmetry of the
energy levels around zero. (The energies are measured relative to the Fermi level.) 
The invariant arclength  and  measure  can be written
using $\delta U_D={\cal O}^T dO$ as
\begin{equation}
\label{eq:classDinfi}
\begin{array}{c}
\displaystyle ds^2_D=\Tr(\delta U_D\delta U_D^T)=2\sum_{k<l} (\delta U_D)_{kl}^2,\\
\quad\\
\displaystyle d\mu_D(U)\propto\prod_{k<l}(\delta U_D)_{kl}.  
\end{array}
\end{equation}
Note that $\delta U_D$ is antisymmetric due to the
orthogonality of ${\cal O}$.  It is seen that $ds^2_D$ [and consequently
$d\mu_D(U)$] is invariant under $\delta U_D\rightarrow W\delta U_DW^T$, with
$W\in$ O$(N)$. 
Such a transformation also preserves the antisymmetry of $\delta U_D$.

The manifold ${\cal M}_{D\rm III}$ is spanned by $U=\tilde{U}\tilde{U}^R$, \mbox{$\tilde{U}\in {\cal M}_D$.} 
It is worthwhile to conjugate with
\begin{equation}
V=\frac{1}{\sqrt{2}}
\left(\begin{array}{cc}
-\sigma_1 & \tau \\
\sigma_1 & \tau
\end{array}\right)
\end{equation}
and define $\tilde{\cal O}=V^T \nu^\dagger \tilde{U} \nu V \in {\rm
  SO}(N)$. The matrix $V$ is chosen such that \mbox{$(\nu V)^\dagger=(\nu V)^R$,}
from which it follows that $\tilde{U}^R=\nu V \tilde{\cal O}^R
V^T\nu^\dagger$. ${\cal M}_{D\rm III}$ is therefore
isomorphic to the manifold spanned by ${\cal O}=\tilde{\cal O}\tilde{\cal O}^R$,
$\tilde{\cal O}\in$ SO$(N)$. 
Defining $\delta \tilde{\cal O}=\tilde{\cal O}^Td\tilde{\cal O}$, the invariant arclength and measure can be written in terms of 
$\delta U_{D{\rm  III}}=\delta \tilde{\cal O}+\delta \tilde{\cal O}^{R}$ as
\begin{equation}\label{eq:classDIIIinfi}
\begin{array}{c}
\displaystyle\!\!\!ds_{D\rm III}^2=\Tr(\delta U_{D{\rm  III}}\ \delta U_{D{\rm  III}}^T)=4
\sum_{k<l}(da_{kl}^2+db_{kl}^2),\\
\quad\\
\displaystyle d\mu_{D\rm III}(U)\propto \prod_{k<l}da_{kl}db_{kl},
\end{array}
\end{equation}
where, in spin grading, 
\begin{equation}
\delta U_{D{\rm  III}}=
\left(\begin{array}{cc}
da &  db \\
db & - da 
\end{array}\right),\quad\!\! da=-da^T,\quad db=-db^T.
\label{eq:deltaKparam}
\end{equation}
The parametrization
\eqref{eq:deltaKparam} follows from \mbox{$\delta U_{D{\rm  III}}=-\delta U_{D{\rm  III}}^T$} and $\delta U_{D{\rm  III}}=
\delta U_{D{\rm  III}}^{R}$. The arclength $ds^2_{D\rm III}$ and the measure $d\mu_{D\rm
  III}(U)$ are invariant under \mbox{$\delta U_{D{\rm  III}}\rightarrow W\delta
U_{D{\rm  III}}W^T$}, with $W\in$ O$(N)$. If $W$ also satisfies $W^{R}=W^{-1}$, 
such a transformation preserves the symmetries of $\delta U_{D{\rm  III}}$.

In the case of the classes $C$ and $C$I we omit the spin degree of freedom,
and we use $N$ to denote the size of the scattering matrices without  spin.
Electron-hole symmetry is now expressed by the relation\cite{Alt97} $U=\Sigma_2 U^*
\Sigma_2$, i.e., $U$ is unitary symplectic. 
 For class $C$ this defines ${\cal
  M}_C= {\rm Sp}(N)$.  The invariant arclength and measure are
\begin{equation}
\label{eq:classCinfi}
\begin{array}{c}
\displaystyle ds_C^2=\Tr (\delta U_C\ \delta U_C^\dagger)=\\
\quad\\
\displaystyle 2\ (\sum_{q=1}^3\sum_{l} (\delta
U_{ll}^{(q)})^2+2\sum_{q=0}^3\sum_{k<l}(\delta
U_{kl}^{(q)})^2),
\end{array}
\end{equation}
\begin{equation}
d\mu_C(U)\propto \left(\prod_{q=1}^3\prod_{l} \delta
U_{ll}^{(q)}\right)\prod_{q=0}^3\prod_{k<l}\delta U_{kl}^{(q)}. 
\label{eq:classCvol}
\end{equation}
Here, $\delta U_C=U^\dagger dU$, with 
\begin{equation}
\delta U_C=\openone^{\rm(eh)}\delta U^{(0)}+ i\sum_{q=1}^3\Sigma_q\delta U^{(q)}, 
\label{eq:quatdecomp}\end{equation}
where $\openone^{\rm(eh)}$ is the identity matrix in electron-hole space, and
$\delta U^{(q)}$ are $N/2\times N/2$ dimensional real matrices. Due to $(\delta
U_C)^\dagger=-\delta U_C$, they satisfy $\delta U^{(0)}=-(\delta U^{(0)})^T$,
and for $q>0$,  $\delta U^{(q)}=(\delta U^{(q)})^T$. The arclength and the
measure are invariant under $\delta U_C\rightarrow W\delta U_CW^\dagger$, with
$W\in {\rm Sp}(N)$. Such a transformation preserves the symmetries of $\delta U_C$ as well. 

The manifold ${\cal M}_{C\rm I}$ is spanned by $U=\tilde U\tilde U^T$, with $\tilde
U\in {\rm  Sp}(N)$. Defining $\delta \tilde U=\tilde U^\dagger d\tilde U$ and decomposing it
according to 
Eq.~\eqref{eq:quatdecomp}, we define  
\begin{equation}
\delta U_{C\rm I}=\delta \tilde U+(\delta \tilde
U)^T=i\Sigma_1\delta \tilde U^{(1)}+
i\Sigma_3\delta \tilde U^{(3)}.
\end{equation}
The invariant arclength and measure are
\begin{equation}\label{eq:classCIinfi}
\begin{array}{c}
\displaystyle ds^2_{C\rm I}=\Tr (\delta U_{C\rm I} \delta U_{C\rm I}^\dagger)=\\
\quad\\
\displaystyle \!\!2\ (\sum_{q=1,3}\sum_{l} (\delta
\tilde U_{ll}^{(q)})^2+2\sum_{q=1,3}\sum_{k<l}(\delta \tilde U_{kl}^{(q)})^2),
\end{array}
\end{equation}
\begin{equation}
d\mu_{C\rm I}(U)\propto \prod_{q=1,3}\prod_{k\leq l}\delta \tilde U_{kl}^{(q)}. 
\label{eq:classCIvol}\end{equation}
The arclength and the measure are invariant under $\delta U_{C\rm I}\rightarrow W
\delta U_{C\rm I} W^\dagger$ with \mbox{$W\in$ Sp$(N)\cap$O$(N)$}. The symmetry of $\delta U_{C\rm I}$ is
also preserved under such a transformation.

\section{Scattering matrix distribution}
\label{sec:sdist}
The scattering matrix $S$ at nonideal coupling can be represented as a combination
of a random $N\times N$ scattering matrix $S_0$ at ideal coupling and a
fixed $2N\times 2N$ scattering matrix $S_c$ responsible for the direct
reflections.\cite{Brouw95gce} 
The matrix $S_0$ is assumed to be uniformly distributed with respect to the
invariant measure on ${\cal M}_X$. The matrix $S_c$ is given by
\begin{equation}
S_c=
\left(\begin{array}{cc}
r &  t' \\
t & r'
\end{array}\right).
\end{equation}
 Here the dimension of all submatrices is $N\times
N$, and all of them carries further structure in electron-hole space, and, for
classes $D$ and $D$III, also in spin space. The matrix $r$ describes direct reflection from the contact, $r'$
describes reflection back to the cavity from the contact and $t$ and $t'$ are
the transmission matrices to and from the cavity, respectively. The scattering
matrices $S_0$ and $S_c$ have the same symmetries. 

The total scattering matrix $S$ is given by
\begin{equation}
S=r+t'S_0(1-r'S_0)^{-1}t, 
\label{eq:S0toS}
\end{equation} 
and the inverse of the relation is
\begin{equation}
S_0=(t')^{-1}(S-r)(1-r^\dagger S)^{-1}t^\dagger.
\label{eq:S0withS}
\end{equation}
We derive the distribution of $S$  from the uniform distribution of
$S_0$ following a similar logic to the calculations in Refs.~\onlinecite{friedman1985aita,gopar2008csd}. 
The starting point of the reasoning is the relation
\begin{equation}
\delta S=M \delta S_0 M^\dagger,
\label{eq:mastereq}
\end{equation}
where $\delta S=S^\dagger dS$, $\delta S_0=S_0^\dagger dS_0$ and
\begin{equation}
M=(1-S^\dagger r)\ t^{-1}.
\end{equation}
The strategy is to express the arclength in an infinitesimal neighborhood
${\cal N}_S$ of $S$ as
\begin{equation}
\!\!ds^2(S)={\rm Tr}(d S^\dagger d S)={\rm Tr}(\delta S^\dagger \delta S)=\sum_{i
  j}g_{ij}(S) dx_idx_j, 
\end{equation}
where $\{dx_i\}$ denotes the set of independent matrix elements of $\delta
U_X$ in the parametrization of an infinitesimal neighborhood ${\cal N}_{S_0}$ of
$S_0$. [${\cal N}_{S}$ is  the  image
of  ${\cal N}_{S_0}$ under the mapping~\eqref{eq:S0toS}.]
This way we can relate the
measure $d\mu(S_0)$  of ${\cal N}_{S_0}$ to the measure $d\mu(S)$ of ${\cal
  N}_{S}$ as\cite{tracy1994irm,RMTQTR}
\begin{equation}
d\mu(S)\propto|\det\ g(S)|^{1/2}\prod_jdx_j\propto|\det\ g(S)|^{1/2}d\mu(S_0),
\end{equation}
where we used that in ${\cal N}_{S_0}$, $d\mu(S_0)\propto\prod_jdx_j$. 
On the other hand, the probability of ${\cal N}_{S}$ is the same as of ${\cal
  N}_{S_0}$, i.e., $P(S)d\mu(S)=d\mu(S_0)$, which gives
\mbox{$P(S)\propto|\det\ g(S)|^{-1/2}$}, 
the distribution we are after.

Parametrizing ${\cal N}_{S}$ with the help of $\delta U_X$ and ${\cal
  N}_{S_0}$ using $(\delta U_X)_0$, the  relation in
Eq.~\eqref{eq:mastereq} can be written as 
\begin{equation}
\delta U_{X}=M'(\delta U_X)_0M'^\dagger
,\end{equation}
where
\begin{subequations}
\begin{align}
M'&=\nu^\dagger M \nu \quad&\textrm{for class  $D$},\\
M'&= V^T\nu^\dagger \tilde{U}^R M (\tilde{U}_0^R)^\dagger\nu V \quad&\textrm{for class $D$III},\\
M'&= \tilde{U}^T M \tilde{U}_0^*\quad&\textrm{for class $C$I}, 
\end{align}
\end{subequations}
and $M'=M$ for class $C$. Here the matrices $\tilde{U}, \tilde{U}_0$ are used
to express $S$ and $S_0$ for class $C$I and $D$III according to
Sec.~\ref{sec:manifolds}, i.e.,  $S=\tilde{U}\tilde{U}^y$,
$S_0=\tilde{U}_0\tilde{U}_0^y$ where $y=T$ and $\tilde{U}, \tilde{U}_0\in {\cal M}_C$ for class
$C$I, and $y=R$ and $\tilde{U}, \tilde{U}_0\in {\cal M}_D$ for class $D$III.
  The matrix $M'$ satisfies
\begin{subequations}\label{eqs:M'props}
\begin{align}
M'&=M'^*\quad &\textrm{for class  $D$},\label{eqs:M'propsD}\\
M'&=M'^*=\tau  M' \tau^T,\   \quad&\textrm{for class $D$III},\label{eqs:M'propsDIII}\\
M'&=\Sigma_2 M'^* \Sigma_2\quad&\textrm{for class $C$},\label{eqs:M'propsC}\\
M'&=M'^*=\Sigma_2 M' \Sigma_2\quad&\textrm{for class $C$I}.\label{eqs:M'propsCI} 
\end{align}
\end{subequations}
The reality of the matrix elements of $M'$ for class $D$ and $D$III follows
from the fact that the set of matrices  satisfying $A=\Sigma_1 A^* \Sigma_1$ is closed
under matrix addition  multiplication and inversion, and that the combination $\nu^\dagger A \nu$ is real. 
We show the proof of $M'=\tau  M' \tau^T$ for class $D$III.   Because of $\nu V=\tau (\nu V)^* \tau^T$, it is
enough to show that $\hat M=\tau \hat M^* \tau^T$, where
\begin{equation}
\hat M=\tilde{U}^R M (\tilde{U}_0^R)^\dagger=\tilde{U}^\dagger S(1-S^\dagger r)t^{-1}S_0^\dagger
\tilde{U}_0.
\end{equation}
It is easy to see that 
\begin{equation}
\tau \hat M^* \tau^T=\tilde{U}^R\ \tau S^*(1-S^T r^*)(t^*)^{-1}\tau^TS_0^R\ (\tilde{U}_0^R)^\dagger.
\label{eq:Mhat}
\end{equation}
Using Eq.~\eqref{eq:S0withS} and the self duality of $S$ and  $S_c$ we find
\begin{equation}
S_0^R=\tau t^*(1-S^T r^*)^{-1}\tau^T S M.
\end{equation}
Substituting in \eqref{eq:Mhat}, and using again the self duality of $S$ leads to
the desired result. The reality of the matrix elements of $M'$ for class $C$I
can be proven following analogous steps. The relation $M'=\Sigma_2 M'^*
\Sigma_2$ for class $C$ and $C$I follows from the closedness of the set of
matrices satisfying $A=\Sigma_2 A^* \Sigma_2$ under matrix addition  multiplication and inversion.

\begin{widetext}
Following from properties \eqref{eqs:M'props}, the matrix $M'$ has a singular value decomposition
\begin{equation}
M'=WDW',
\label{eq:M'SVD}\end{equation}
where
\begin{subequations}
\begin{align}
&\quad D={\rm diag}(d_k), \qquad\qquad\ \quad  k=1\dots N  \ &\textrm{for class $D$,}\\
&\quad D={\rm diag}(d_k)\ \openone^{\rm (sp)},   \quad\qquad k=1\dots \frac{N}{2} &\textrm{for class   $D$III,}\\
&\quad D={\rm diag}(d_k)\ \openone^{\rm (eh)},   \quad\qquad k=1\dots \frac{N}{2} &\textrm{for class   $C$,}\\
&\quad D={\rm diag}(d_k)\ \openone^{\rm (eh)},  \quad\qquad k=1\dots \frac{N}{2}&\textrm{for class   $C$I,}
\end{align}
\end{subequations}
with $\openone^{\rm (sp)}$ being the identity  matrix in spin space, and
\begin{subequations}
\begin{align}
&\quad W, W'\in {\rm O}(N)\quad &\textrm{for class $D$,}\\
&\quad W=(W^R)^{-1}, W'=(W'^R)^{-1}\in {\rm O}(N)\quad  &\textrm{for class   $D$III,}\\
&\quad W, W'\in {\rm Sp}(N)\quad &\textrm{for class $C$,}\\
&\quad W, W'\in {\rm Sp}(N)\cap {\rm O}(N)\quad &\textrm{for class $C$I.}
\end{align}
\label{eq:Wsymm}
\end{subequations}
Using the decomposition \eqref{eq:M'SVD}, the invariant arclength reads 
\begin{equation}
ds^2(S)=\Tr(\delta U_{X}\delta U_{X}^\dagger)=\Tr\left\{[D(\delta U'_X)_{0}D][D(\delta U'_X)_{0}D]^\dagger\right\},
\end{equation}
where we used the parametrization  $(\delta U_X)_0= W'^\dagger(\delta U'_X)_{0}W'$. 
From the  properties of $W'$ in Eq.~\eqref{eq:Wsymm} it follows that the matrix $(\delta U'_X)_0$ has the same
symmetries as $(\delta U_X)_0$. It is easily read off that
\begin{subequations}
\begin{align}
&\qquad\quad \sqrt{\det\ g(S)}\propto\prod_{k<l}^N d_kd_l=\prod_k^N d_k^{N-1}=|\det M|^{N-1} &\textrm{for class $D$,}\\
&\qquad\quad \sqrt{\det\ g(S)}\propto\prod_{k<l}^{N/2} (d_kd_l)^2=\prod_k^{N/2} d_k^{N-2}=|\det M|^{\frac{N}{2}-1} &\textrm{for class $D$III,}\\
&\qquad\quad \sqrt{\det\ g(S)}\propto\prod_{k<l}^{N/2}
(d_kd_l)^4\ \prod_j^{N/2}d_j^6=\prod_k^{N/2} d_k^{2N+2}=|\det M|^{N+1}  &\textrm{for class $C$,}\\
&\qquad\quad \sqrt{\det\ g(S)}\propto\prod_{k\leq l}^{N/2}
(d_kd_l)^2=\prod_k^{N/2} d_k^{N+2}=|\det M|^{\frac{N}{2}+1} &\textrm{for class $C$I.}
\end{align}
\end{subequations}
\end{widetext}
The distribution of $S$ is therefore given by
\begin{equation}
P(S)\propto |\det(1-r^\dagger S)|^{-(N/t+\sigma)},
\label{eq:P(S)}\end{equation}
where $t=1$ in the absence of time reversal invariance and $t=2$ otherwise,
and   $\sigma=-1$ in the absence of spin rotation invariance and $\sigma=1$ otherwise.

\section{Conductance distribution for an Andreev quantum dot in a magnetic field.}
\label{sec:appliC}
\begin{figure}
\begin{center}
\includegraphics[height=6cm]{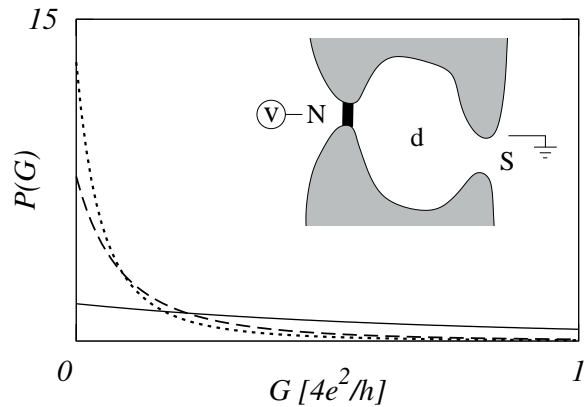}
\end{center}
\caption{Conductance distribution for a single mode chaotic Andreev quantum dot in
  a magnetic field for different values of the barrier transparency $\Gamma$. 
A sketch of the system
 is shown in the inset. 
 The barrier is indicated by a black rectangle.
The solid,  dashed, and dotted curves correspond to $\Gamma=0.9$, $\Gamma=0.6$, and $\Gamma=0.5$,
  respectively. 
   }
\label{fig:classCcond}
\end{figure}

To illustrate the use of our result, we calculate the conductance distribution
for  a chaotic Andreev quantum dot in a magnetic field. 
We assume that the spin-orbit
scattering is negligible, i.e., the system belongs to symmetry class $C$. 
For simplicity, we consider $N=2$, which is the minimal dimension of $S$ due to the electron-hole structure. 
This corresponds to the case that the quantum dot is connected to a normal
reservoir via a single mode point contact. A sketch of the system is shown in
the inset of  
Fig.~\ref{fig:classCcond}. The point contact is assumed to
contain a tunnel barrier of transparency $\Gamma$. 
The  barrier alone does not mix electrons and holes, therefore its reflection
matrix is diagonal in electron-hole space,
\begin{equation}
r=\sqrt{1-\Gamma}
\left(\begin{array}{cc}
e^{i\xi} & 0\\
0 & e^{-i\xi}
\end{array}\right)=\sqrt{1-\Gamma}\ {\rm exp}(i\xi \Sigma_3).
\end{equation}
Here $\xi$ is the phase an electron acquires upon reflection from the
barrier. 
The total scattering matrix $S$ is distributed according to $P(S)$ in the group
\mbox{Sp$(2)\equiv$ SU$(2)$}. The conductance in units of $4e^2/h$ is given by\cite{takane1991,lambert1991glf,bee92a}
\begin{equation}
G(S)=|S_{\rm he}|^2.
\end{equation}
Writing the total scattering matrix as $S={\rm exp}(i\xi \Sigma_3)\ U$, $U\in
SU(2)$, and using that $|S_{\rm he}|^2=|U_{\rm he}|^2$ and \mbox{$d\mu_C(S)=d\mu_C(U)$}, the conductance
distribution is given by
\begin{equation}
P(G)=\int_{{\rm SU}(2)}\frac{\delta(G-|U_{\rm he}|^2)}{|\det(1-\sqrt{1-\Gamma}U)|^3}d\mu_C(U).
\end{equation}
\ \\ 
Using the Euler angle parameterization for SU(2),
\begin{equation}\begin{array}{c}
U\!\!=\!\!\!\!
\ 
\left(\begin{array}{cc}
e^{-i(\phi+\psi)/2}\cos(\theta/2) &-e^{i(\psi-\phi)/2}\sin(\theta/2)\\
e^{i(\phi-\psi)/2}\sin(\theta/2)&e^{i(\phi+\psi)/2}\cos(\theta/2)
\end{array}\right)\!\!,\\
\quad \\
\quad \\
(\phi,\psi,\theta) \in [0,2\pi]\times [0,4 \pi]\times [0,\pi]\equiv {\cal D},
\end{array}\end{equation}
the measure is $d\mu_C\propto\sin(\theta)$, and
$|U_{\rm he}|^2=\sin^2(\theta/2)$. The integral
\begin{equation}
\begin{array}{c}
\displaystyle P(G)=\frac{\Gamma^3}{16\pi^2}\int_{\cal D} d\theta d\phi d\psi
F_\Gamma(\theta,\phi,\psi)\\
\quad\\
\displaystyle F_\Gamma(\theta,\phi,\psi)=
\frac{\sin(\theta)\delta[G-\sin^2(\theta/2)]}{[2-\Gamma-2\sqrt{1-\Gamma}\cos{\frac{\phi+\psi}{2}\cos\frac{\theta}{2}}]^3}
\end{array}
\end{equation}
can be evaluated in closed form, resulting in 
\begin{equation}
P(G)=\Gamma^3\ \frac{\Gamma^2+2(G-3)(\Gamma-1)}{[\Gamma^2-4G(\Gamma-1)]^{5/2}}
\end{equation}
for $0\leq G\leq 1$, and $0$ otherwise.
In Fig.~\ref{fig:classCcond} we show $P(G)$ for different values of the barrier
transparency. It is seen that the uniform distribution $P(G)=1$ corresponding
to ideal coupling ($\Gamma=1$) is gradually transformed into a distribution
that is peaked at $G=0$ as the transparency decreases. The first two moments of the conductance
are given by  
\begin{equation}
\langle G \rangle=\frac{\Gamma^2}{2},\quad \langle G^2\rangle=\frac{\Gamma^3}{3}.
\end{equation}

\section{Conclusion}
\label{sec:concl}
In conclusion, we have calculated the distribution $P(S)$ of the scattering
matrix  at the Fermi energy for  chaotic Andreev quantum dots in the
nonstandard symmetry classes of Altland and Zirnbauer. Our result, which 
allows for arbitrary coupling to the transport channels, is based on the assumption that the
scattering matrix is uniformly distributed in ${\cal M}_X$ for the case of
ideal coupling, i.e., in the absence of direct
reflections from the openings of the Andreev quantum dot.

Apart from the symmetry class dependent exponent, our result $P(S)$ has a similar structure to
the Poisson kernel distribution $P_{\beta}(S)$  corresponding to Dyson's
standard symmetry classes. As a closing remark, we would like to emphasize an aspect
in which $P(S)$ and $P_\beta(S)$ are different. $P_{\beta}(S)$
can be obtained\cite{mello1985ita} as the distribution of 
unitary matrices (with the symmetry corresponding to $\beta$)  that maximizes the information entropy,  subject to the constraint 
 $\langle S^p\rangle=\langle S\rangle^p\equiv r^p$, where $r$ is a subunitary
matrix. This analyticity-ergodicity
constraint follows\cite{mello1985ita} from  the requirement that the scattering
matrix has poles only in the lower half of the complex-energy plane (analyticity),
and the assumption that spectral averages equal  ensemble averages (ergodicity).
Given the similar form of $P(S)$ and $P_\beta(S)$, one might wonder whether
$P(S)$ can be obtained by the same maximization procedure. As we show below,
the answer is negative. 
In the presence of superconductivity, spectral average can mean two types of
averages. First, it can refer to averaging scattering matrices over an
interval of excitation energies $\varepsilon$.  
Since electron hole-symmetry relates scattering matrices at $\varepsilon$ and
$-\varepsilon$, it results in the additional constraint $S=\Sigma_1 S^*
\Sigma_1$ only at $\varepsilon=0$.
Therefore, such an average would be over scattering matrices with different
symmetry than the matrices in the ensemble corresponding to $\varepsilon=0$,
which violates the ergodicity assumption.
 Second, the
 spectral average can refer to an average over an interval of
Fermi energies of the superconductor, while the excitation energy is kept at
$\varepsilon=0$. 
 However, if ${\cal E}$ is a pole of the
scattering matrix as the function of the Fermi energy, so is ${\cal
  E}^*$. To see this, one turns to the channel coupled model used in Ref.~\onlinecite{Alt97},
in which the poles of the scattering matrix on the complex-Fermi energy plane are
eigenvalues of a matrix $({\cal H}-iWW^\dagger)\Sigma_3$, where ${\cal H}$ models
the Bogoliubov de Gennes Hamiltonian (at a fixed Fermi energy) and $W$ is a
coupling matrix. The conjugation  relation between the poles is the consequence of $\Sigma_1 ({\cal H}-iWW^\dagger)^*
\Sigma_1=-({\cal H}-iWW^\dagger)$.  This
precludes the use of  the analyticity properties as in
Ref.~\onlinecite{mello1985ita}. 
We thus find that the  analyticity-ergodicity constraint is
not applicable to the ensembles studied in this paper. This is already
signaled in the case of ideal coupling. For example for class $D$, one
has\cite{Alt97} $\langle S \rangle=0$, but $\langle \Tr S^2 \rangle=1$. For
the more general, nonideal coupling case described by $P(S)$, one also finds
that $r\neq \langle S \rangle$, in contrast to the case of
the Poisson kernel $P_{\beta}(S)$. This can be illustrated on the example
in Sec.~\ref{sec:appliC}, where, for $\xi=0$,  $\langle \Tr S
\rangle=\sqrt{1-\Gamma}(\Gamma+2)$, as opposed to $\Tr\ r=2\sqrt{1-\Gamma}$.

\section*{ACKNOWLEDGMENTS}
I thank C.~W.~J.~Beenakker for valuable discussions. 
This work was supported by the Dutch Science Foundation NWO/FOM.

\end{document}